\documentclass[acmsmall, screen]{acmart}
\AtBeginDocument{%
  \providecommand\BibTeX{{%
    \normalfont B\kern-0.5em{\scshape i\kern-0.25em b}\kern-0.8em\TeX}}}

\usepackage{exii-macros}

\usepackage{booktabs}

\usepackage{multirow}
\usepackage{tabularray}

\usepackage[commandnameprefix=always, final]{changes}

\setcopyright{acmcopyright}
\copyrightyear{2018}
\acmYear{2018}
\acmDOI{XXXXXXX.XXXXXXX}

\acmConference[CSCW '25]{Make sure to enter the correct
  conference title from your rights confirmation emai}{Oct 18--22,
  2025}{Bergen, Norway}
\acmBooktitle{Woodstock '18: ACM Symposium on Neural Gaze Detection,
 June 03--05, 2018, Woodstock, NY} 
\acmPrice{15.00}
\acmISBN{978-1-4503-XXXX-X/18/06}

\begin{document}

\title{Understanding Young People’s Creative Goals with Augmented Reality}

\author{Amna Liaqat}
\affiliation{%
  \institution{Princeton University}
  \country{USA}
}
\email{al0910@princeton.edu}

\author{Fannie Liu}
\affiliation{%
  \institution{JPMorgan Chase \& Co}
   \country{USA}
  }
\email{lfannie.liu@jpmchase.com}

\author{Brian Berengard}
\affiliation{%
  \institution{The Clubhouse Network}
   \country{Argentina}
  }
\email{berengard.brian@gmail.com}

\author{Jiaxun Cao}
\affiliation{%
 \institution{Duke Kunshan University}
  \country{China}
 }
 \email{jc851@duke.edu}

\author{Andrés Monroy-Hernández}
\affiliation{%
  \institution{Princeton University}
   \country{USA}
  }
  \email{andresmh@cs.princeton.edu}

\renewcommand{\shortauthors}{Liaqat et al.}

\begin{abstract}
 Young people are major consumers of Augmented Reality (AR) tools like Pokémon GO, but they rarely engage in creating these experiences. \chadded{Creating with technology gives young people a platform for expressing themselves and making social connections. However, we do not know what young people want to create with AR, as existing AR authoring tools are largely designed for adults.} \chdeleted{The ability to create with today's technology can enable people to more actively participate in society, yet we do not know what young people would like to create to express themselves.} To investigate \chreplaced{the requirements for an AR authoring tool} {how to support creative character-based self-expression in AR}, we ran eight design workshops with 17 young people in Argentina and the United States that centered on young people's perspectives and experiences. We identified four ways in which young people want to \chadded{create with} AR, \chadded{and contribute the following design implications for designers of AR authoring tools for young people}: (1) \chreplaced{Blending imagination into AR scenarios to preserve narratives}{leveraging location, time to design AR interactions with the physical world}, (2) \chreplaced{Making traces of actions visible to foster social presence}{making asynchronous multiplayer games that preserve contribution through visible digital traces} (3) \chreplaced{Exploring how AR artifacts can serve as invitations to connect with others}{respecting autonomy when sharing creations through mimicking of social conventions in AR}, and (4) \chreplaced{Leveraging information asymmetry to encourage learning about the physical world}{leveraging information asymmetry as learning opportunities}. \chdeleted{From these, we contribute four implications for designers to support young people in becoming creators of AR technology.}  
\end{abstract}

\ccsdesc[500]{Human-centered computing~Collaborative and social computing}

\keywords{augmented reality, design workshops, young people, creativity}


\received{16 January 2024}
\received[revised]{16 July 2024}
\received[accepted]{20 September 2024}

\begin{teaserfigure}
  \includegraphics[width=\textwidth]{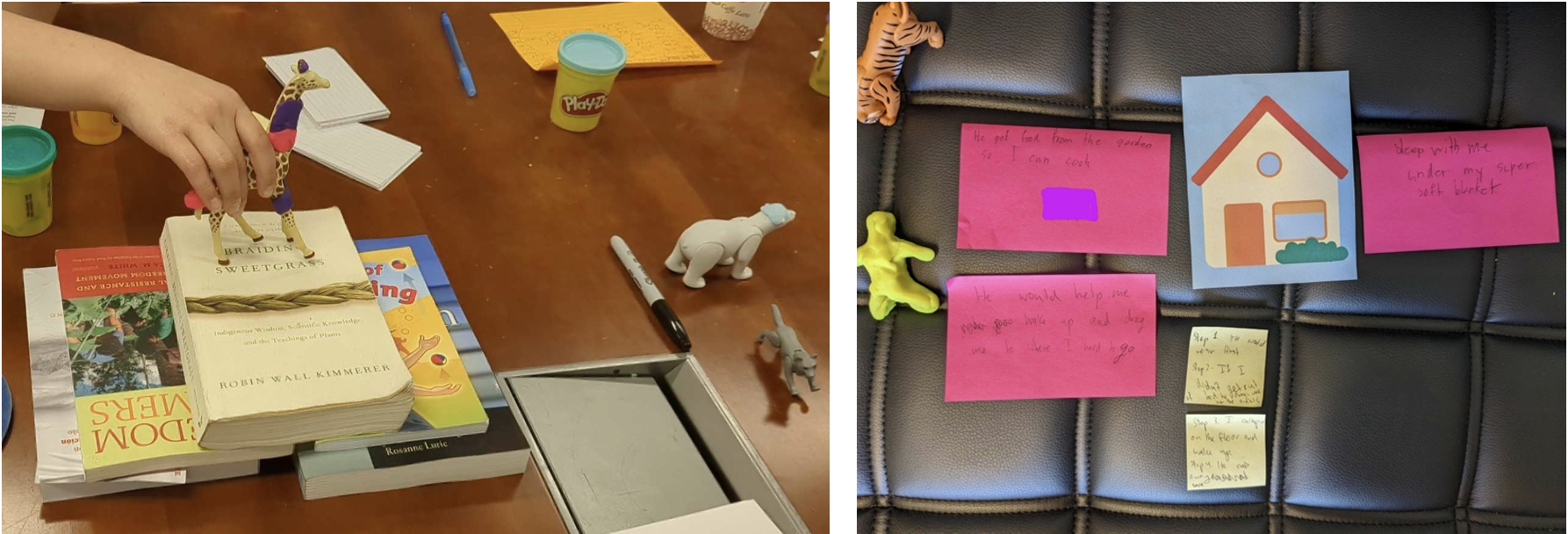}
    \caption{Examples of the physical design activity. A participant manipulates a plastic giraffe, with access to playdough and sticky notes (left). An example of sticky notes participants have placed to detail what happens at ``home'' with their plastic tiger (right).} 
    \label{fig:setup}
\end{teaserfigure}

\maketitle

\section{Introduction}

Young people (preteens and teenagers) today are frequently engaged in Augmented Reality (AR) activities~\cite{radu2009augmented, das2017augmented, paavilainen2017pokemon, hnatyuk23augmented}. Many young people play AR games with location-based digital characters on Pokémon GO \cite{PKMNGO, paavilainen2017pokemon}, they adorn their videos with AR effects on TikTok \cite{TikTok}, or they converse using face-altering AR filters on Snapchat \cite{Snapchat}. However, young people are merely given those AR experiences specifically for them to consume. They rarely engage in \textit{creating} those AR experiences, leaving young people without a voice in shaping the future of AR experiences despite being major users ~\cite{hnatyuk23augmented}.\chadded{ Though young people can articulate complex ideas on what they think about popular technologies, they are often not included as stakeholders when developing technology \cite{solyst2023potential}. Motivated by this oversight, we were interested in involving young people as stakeholders and hearing their ideas on what they would do with an AR authoring tool designed for them.} Although HCI researchers have tried to address this by developing tools for novices to more easily create AR experiences \cite{blocksGuo, lunding2022exposar}, there is little research aimed at \textit{understanding} what young people would like to \textit{create} with AR\chadded{, and how they would like to connect with other younger people in the process of creation.} 

\chdeleted{For people - including the younger members of our communities - to be active participants in society, they should be able to create with today's digital technology \cite{resnick_pianos_1996, resnick2008sowing}.} 

Since the early days of computing, young people have enjoyed playing with digital characters, such as the turtle from Logo \cite{solomon1978teaching}, various animals in Moose Crossing \cite{bruckman1997moose}, and sprites from Scratch \cite{resnick2009scratch}. More recently, the success of Pokémon GO reaffirmed young people's interest in interacting with and controlling digital characters. Additionally, studies of desktop-based tools show that young people enjoy the freedom to express themselves in digital settings through creating scenarios such as animating characters, creating stories, and interacting with other users ~\cite{resnick_pianos_1996, bruckman1997moose, scratch_foundation_scratch}. Expanding on this, we investigate how to leverage AR's close ties between the digital and the physical to identify how young people can express themselves through character-based immersive experiences that connect with the world and people around them. We focus our investigation on character-based scenarios as it is an established medium through which young people enjoy interacting in digital spaces \cite{solomon1978teaching, bruckman1997moose}. 
More specifically, we aimed to answer the following guiding question:

\begin{itemize}
    \item \textbf{RQ: }What types of playful, character-based experiences do young people envision they could craft with \chadded{an AR authoring tool}?
\end{itemize}

To answer our guiding question, we ran eight one-hour workshops with 17 children split into groups of 2 to 3 participants. We recruited participants aged 11--16 in the United States and Argentina. Participants completed a physical and digital design activity where they manipulated characters to brainstorm scenarios they would like to create. We recorded these sessions and analyzed them through thematic analysis. 

From our analysis, we derived the following findings, discovered \chadded{underlying goals and values}, and identified design opportunities for designers of AR \chreplaced{authoring tools}{technology}. We found that young people:

\begin{enumerate}
    \item \chreplaced{Explored different ways to immerse their AR characters in the physical world using
technical features and their imaginations.}{Want to realistically ground AR characters in the physical world, but they can use their imaginations to craft narratives that overcome the limitations of AR technology.}
    \item Created asynchronous scenarios anchored by location, but it is important to them to leave traces that make their contributions visible.
    \item \chreplaced{Fostered a sense of togetherness and friendship between AR characters through meetings in physical space.}{Are curious about other young peoples' creations, but this curiosity must be balanced with respecting autonomy. This balance can be achieved by leveraging AR's spatial "discovery" to mimic social conventions.}
    \item \chreplaced{Wanted information asymmetry between characters and users to create learning opportunities.}{Are interested in leveraging information asymmetry to craft social interactions, so young people should be able to use AR artifacts to discover and gather information about the world that is interesting to them.} 
\end{enumerate}

Hence, the contributions of this work to the CSCW community are two-fold. First, through running design workshops, we provide empirical insights into young people's creative and \chadded{social} goals \chadded{when creating with AR}. Second, we propose design opportunities for \chreplaced{for build AR authoring tools that support the goals of young people}.

\section{Related Work}
To understand the potential of AR to enable young people to create, we present studies on non-AR tools that have supported young people in creating with technology. \chadded{Next, as most AR authoring tools are designed for adults, we describe what kinds of experiences adults have had with these tools. We then explain the importance of involving young people in designing technology for them.} Despite the unique and valuable perspectives young people can bring to the design table, they are often not included in the technology development process until the end stages. We discuss why a user-centered approach is especially valuable as HCI researchers investigate how we can leverage the relatively new medium of AR to enable young people to create rather than consume. \chadded{Finally, we present prior work on how young people have used AR tools, and we discuss how we still need to understand what young people want to create with AR.}

\subsection{\chreplaced{What have young people created with non-AR authoring Tools?}{Young people as creators of technology}}

Creating provides children with an avenue for self-expression. Room for self-expression is important for children to grow into active members of society~\cite{resnick_pianos_1996}. Through the expression of their personalities, children learn to communicate their identity to their peers and build social relationships  \cite{tomova_importance_2021}.  This process of navigating their evolving identities as they interact with peers helps children develop a social sense of self \cite{erikson_identity_1994}.  

Through the different epochs of computing, researchers have developed platforms to enable children and adolescents to creatively build and self-express in digital forms.  As early as the 1960s, researchers created the Logo programming environment, which enabled children to manipulate a digital turtle's movements to paint geometric figures and to problem solve, verbalize abstract ideas, create metaphors, and engage with mathematical concepts \cite{solomon1978teaching}. Years later, with the advent of massively parallel computing in the late 1990s, StarLogo enabled children to use blocks-based programming to control many turtles at once to create simulations of complex systems \cite{colella2001adventures, resnick1996starlogo}.

Around the same time, in 1997, researchers created a social computing platform called Moose Crossing where children could collectively construct a virtual online world and program characters that interacted with the world they had built and with other participants \cite{bruckman1997moose}. More recently, in 2007, Scratch was created to empower children to create web-based animated stories, interactive art, and video games, and collaborate with others in building them \cite{resnick2009scratch}. 
These investigations demonstrated how children can creatively express themselves in digital spaces. Moose Crossing and Scratch additionally illustrated how social collaboration plays a key role in creating engaging virtual experiences. Both Moose Crossing and Scratch leveraged digital media as tools for expressiveness by providing a toolkit of manipulation options through which young people can play and learn. In the present work, we ask: how will these ideas manifest in spatial computing, such as AR? What would young people want to make in this new computational canvas? 

There are AR-based creativity support tools that enable young people to make interactive experiences and see their creations in AR in their physical space ~\cite{radu2009augmented, lunding2022exposar}, and to manipulate smart objects (\eg{}, smart lights, fans, thermostats) in an Intelligent Living room ~\cite{stefanidi2021children}. AR has additionally been used to support storytelling through co-located embodied playful interactions \cite{dagan2022project}. However, while these works 
allowed some room for self-expression, they were limited in that the young people in these studies did not provide input on defining scenarios. Participants were presented with AR prototypes that had clear objectives established by the research team ~\cite{radu2009augmented, dagan2022project, stefanidi2021children} or other stakeholders \cite{lunding2022exposar}. Therefore, while participants did demonstrate the potential to create and express themselves in AR within an externally defined context, we lack a holistic understanding of firstly, how the affordances of AR can support creative expression and, secondly, how young people would like to use AR in ways that are engaging to them.


Desktop-based applications such as Moose Crossing and Scratch have demonstrated the breadth and depth of creative production children are capable of when given the flexibility to create. Expanding on these prior works, we aim to similarly identify young people's capabilities and desires to create with AR and its unique affordances of overlaying virtual content in physical space. \chadded{In the next section, we present what adult users have created with AR authoring tools, and discuss what these tools may suggest about understanding young people's needs in creating with AR.} 

\subsection{\chadded{What do adults create with AR authoring tools?}}
\chadded{
There are many AR authoring tools for adults that support users, both novices and experts, in creating experiences with AR. Many of these are commercial products where users can drag and drop virtual objects to create scenes (e.g., Reality Composer) and add interactivity to AR elements (e.g., Adobe Aero and Rapido \cite{leiva2021rapido}). With these tools, users have created a variety of experiences, such as recording video and overlaying it with animated AR elements, creating realistic physics, and making stories. On the commercial side, social media platforms offer AR tools for developers. With Meta Spark and Snap AR, users with coding experience have built complex AR experiences such as AR filters, multiplayer games, and educational tools. While these tools require coding knowledge, other AR authoring tools are intended for rapid prototyping and prioritize ease of use for novices. For instance, users with no coding or 3D modeling experience have sketched elements and added interactivity using rapid prototyping tools like PintAR, which is a head-mounted system, and Pronto, a mobile app \cite{gasques2019pintar, leiva2020pronto}. While all of these authoring tools vary in complexity of use, they all aim to enable users to create realistic AR experiences. In these tools, users have achieved realism by bringing physical objects into a digital space, adding animations to sketches, and incorporating realistic physics into AR objects. Though adults have created diverse experiences with AR authoring tools, we have limited understanding around what kinds of scenarios young people are interested in making. In the following section, we explore the benefits of adopting user-centered approaches when investigating young people's technology preferences.    
 }

\subsection{\chreplaced{Why is it important to center young people in the design process?}{Centering Young People in the Design Process}}

Though young people are enthusiastic users of technology and bring unique perspectives, they are often excluded until the final stages of technology development \cite{druin_role_2002}. Leaving young people out of the design process means that their experiences, values, and goals are not reflected in the tools developed by researchers \cite{mcnally2016children}. Involving young people in the design process increases their sense of ownership of the resulting solutions and sense of pride over their contributions \cite{liaqat2021participatory}. 

Some HCI research has employed user-centered approaches to address these issues when working with under-served populations \cite{vines2012cheque}. These user-centered approaches may be co-design, participatory design (PD), or variants of these methods \cite{vines2015beginnings}. These approaches are intended to elicit participants’ voices through hands-on design workshops and address power imbalances between researchers and participants by giving young people an active role in creation. For example, Yip and colleagues emphasized the importance of meaningful partnerships across the entire co-design cycle to mitigate the imbalances between an adult researcher and a child participant \cite{yip2017examining}. These participatory approaches have been employed successfully to empower children’s voices in the design and evaluation process to design a wide array of tools ranging from storytelling robots to digital libraries to smart objects \cite{druin2009cooperative, frauenberger2019thinking}. 

The rise of popularity in AR has led to many emerging applications that aim to offer engaging experiences for young people. However, young people generally were not involved in the early stages of designing these experiences. Outside of AR, researchers have successfully developed fun and useful applications by engaging young people in the design process \chadded{through co-design techniques}, such as a piano practice application for young people (ages 11-17) \cite{birch2023diverse}, an educational computer game for teenagers with special learning needs \cite{danielsson2006participatory}, and an app for cooperative storytelling with children aged 10-11 \cite{garzotto2008broadening}. Drawing inspiration from these prior works, we explore the design of AR experiences for young people by similarly including them in the design process. In particular, we highlight the potential for young people to explore new kinds of interactions in AR, as AR provides a unique opportunity to integrate the physical and digital worlds in engaging and playful ways~\cite{zund2015augmented, dagan2022project}. Identifying what these interactions may look like requires validating the experiences and knowledge of young people, who are some of the primary early adopters of AR due to its wide adoption on social platforms used by youth \cite{hnatyuk23augmented}. 

Thus, in this work, we ran workshops to generate a collection of design inspiration for AR researchers that reflects the interests, expectations, and voices of young people. In the next section, we present work that has been done on identifying what young people want to create with AR and what remains to be explored about young people's needs in an AR authoring tool.

\subsection{\chadded{What are young people's experiences with AR?}}

\chadded{Some prior work has investigated how young people have used AR tools designed for them in educational contexts. A study of 46 teenagers aged 13-15 found that when children can bring elements from their own world (e.g., scanning an object from their environment) into the AR world, it promotes engagement and reflection on the learning activity \cite{lunding2022exposar}. They found that teenagers enjoyed experimenting with location in AR to manipulate user behavior, for example, by creating a path in AR that takes the user to a specific store while intentionally avoiding another store \cite{lunding2022exposar}. Another study with nine children aged 5-12 looked into how children combined paper and digital elements to create interactive AR sprites and control them using hand gestures to learn computational thinking \cite{im2021draw2code}. They found that when children can personalize their creations (e.g., adding their drawing into the AR world), it encourages them to explore further, exposing them to basic programming concepts through playful and tangible interaction \cite{im2021draw2code}. Children's unique interactions with AR have been explored in a mental model study. The study involved 65 children between 8-13 years and found that physical manipulation with AR, in the form of spatially orienting handmade paper cubes, can promote engagement with an activity as tangible interactions leverage children's spatial mental models \cite{fuste2019hypercubes}. Thus, the study suggests further exploration into the connection between children's mental models and the physical affordances of AR \cite{fuste2019hypercubes}.}

\chadded{
AR also opens possibilities for unique models of collaboration as its grounding in physical space encourages co-located interactions. For instance, in a study where nine child-parent dyads added AR animation to paper drawings to explore computational thinking concepts, AR was found to encourage working in parallel as users can engage in physical activity side-by-side while still collaborating with one another \cite{im2021draw2code}. Leveraging a group's shared attention is another advantage of collaboration in AR, as non-verbal cues are observable, as was found in the AR and VR tool, MR. Brick, where 24 children aged 6-10 worked together to assemble blocks \cite{wu2023mr}. Collaboration in AR was also explored through the tool, StoryMakAR. Nineteen teenagers aged 14-18 customized IoT devices and enthusiastically collaborated with their peers to make a story to go along with their creations, and learned programming skills in the process \cite{glenn2020storymakar}. These studies on collaboration in AR provided fixed tasks for children to complete. Co-creation in AR may present new opportunities for collaboration when young people can author their own experiences. We suggest that there are open opportunities to explore how children use shared attention and non-verbal cues to cooperate and collaborate in AR and perhaps discover novel ways in which children can collaborate in creation activities. 
}


\chadded{
The studies described in this section show that young people have used and benefited from AR technologies in interesting ways, such as through creating games, collaborating with peers, and learning something new. However, this prior work has investigated \textit{how} young people use AR when given a tool. What remains to be explored is \textit{what} young people would like to make with an an AR authoring tool. In prior work on understanding what young people create with AR, the purpose of the systems has frequently been educational, often to teach computational thinking. In contrast, our purpose is to expose the broader goals young people have when creating with AR beyond learning, such as their creative interests, ways of self-expressing, and socialization with their peers. In this work, we present a series of design workshops that center the voices of young people and investigate what AR experiences could look like when young people have the freedom to create.    
}

\section{Methods}
To identify what young people \chreplaced {want from an AR authoring tool,} {need to express themselves through AR,} we ran eight design workshops with 17 young people. We set the age range for young people between 11-16 (pre-teens to mid-teens), \chadded{motivated by design workshops that effectively worked with a similar age range \cite{birch2023diverse, danielsson2006participatory}.} \chdeleted{Since Generation Z frequently use AR, we selected this age range to encompass members of Generation Z who are still children.} In these workshops, \chadded{we aimed to foster participants' imaginations through playful physical and digital design activities }where participants brainstormed AR experiences they would like to create. \chadded{We informed our design activities from co-design studies with children, where low-fidelity activities that included paper, markers, and open-ended prompts encouraged participants to brainstorm novel ideas that were not limited by technical feasibility \cite{liaqat2022hint, xie2012connecting}} \chdeleted{Workshops consisted of both physical and digital design activities.} In designing our workshops, we took inspiration from Logo, where children were encouraged to express themselves through programming an anthropomorphized turtle, which helped prompt problem-solving, verbalizing abstract ideas, and creating metaphors \cite{solomon1978teaching}. In this work, we aimed to elicit similar self-expression from young people. We asked participants to play with a character and think about scenarios and interactions the character could have with the physical world, with other characters, and with other participants. Through these activities, we aimed to elicit participants' preferences, goals, and opinions on what experiences they would like to create with AR.

\subsection{Participants}
We recruited participants aged 11-16 from the United States and Argentina. We ran studies in two countries to broaden the range of experiences and voices in our workshops.  Participants in the workshop knew each other - as friends, siblings, or classmates. Participants' average age was 13.7 (\textit{SD}=1.8). Eight participants identified as female, and nine as male. 
The detailed demographic information of participants is provided in Table~\ref{tab:demographic}. \chadded{We also asked participants to self-report their usage of social platforms that have AR features. Fifteen of the 17 participants (excluding P6 and P7) had used at least one of the following apps with AR features: Snapchat lens/filters, TikTok effects, Instagram filters, and Pokemon Go. Sixteen of 17 participants (excluding P15) had programming experience with block-based apps like Scratch, Blockly, Tynker, and Android App Inventor.} US participants resided close to the university where we held the workshops. We recruited participants in the US via flyers that were aimed at parents and snowballing. We recruited Argentinian participants with the assistance of a non-profit organization that provides digital literacy classes in Buenos Aires, where one of the authors worked. We compensated US participants with a \$15 Amazon gift card. After consultation with the local non-profit, we elected not to provide monetary compensation to Argentinian participants to avoid coercion. \chadded{To ensure that there were no major differences between the two groups, we ran our data analysis on each group (US and Argentinian Participants) independently. We did this by comparing the ideas generated between the two groups and looking for any unique themes that may be present in one group but not the other. As we identified no major differences in the results, we proceeded with combining the groups for our analysis.} The University IRB approved the study. \chadded{All participants signed an assent form. Parents also signed a consent form for their children. Before signing, the researcher verbally went over the form with the participants, emphasizing that their participation was voluntary, that they could take a break or withdraw at any time, and that all data would be anonymized.}

  \begin{table*}[h]
  \begin{tabular}{p{2cm}p{2cm}p{1cm}p{1.5cm}p{3cm}p{2cm}}
    \toprule
    \textbf{Workshop ID} & \textbf{Participant ID} & \textbf{Age} & \textbf{Gender} & \textbf{Coding Tools} \\
    \midrule
    US1 & P1US1 & 11 & Female & Scratch, Tynker  \\
        & P2US1 & 11 & Female & Scratch, Hour of Code, Programming Robots, Replit (Python) \\
        & P3US1 & 11 & Female & None \\
    \midrule
    US2 & P4US2 & 14 & Male & Scratch \\
        & P5US2 & 14 & Male & Scratch, Tynker \\  
    \midrule
    US3 & P6US3 & 11 & Female & Scratch, Blockly\\
        & P7US3 & 12 & Female & Scratch, Blockly, Hour of Code\\
    \midrule
    US4 & P8US4 & 15 & Male & Scratch, Blockly, Tynker, Arduino, Khan Academy Code\\
        & P9US4 & 14 & Male & Blockly \\
    \midrule
    AG1 & P10AG1 & 15 & Female & PSeInt \\
        & P11AG1 & 13 & Male\\
    \midrule
    AG2 & P12AG2 & 15 & Female & Scratch\\
        & P13AG2 & 15 & Male & Arduino\\  
    \midrule
    AG3 & P14AG3 & 15 & Male & Scratch, Android App Inventor, Tynker\\
        & P15AG3 & 15 & Female & None\\
    \midrule
    AG4 & P16AG4 & 16 & Male & Scratch\\
        & P17AG4 & 16 & Male & LUA\\
 \bottomrule
  \end{tabular}
  \caption{Demographic information of participants. US1 represents the first workshop conducted in the US, while AR1 represents the first workshop conducted in Argentina.}
  \label{tab:demographic}
\end{table*}

\subsection{Design Workshops}
Design workshops took place in a room on a university campus in the United States, and at the non-profit organization in Argentina. Workshops lasted approximately one hour. Participants \chadded{and their parents} signed consent forms and completed a demographic questionnaire. \chadded{Parents were invited to stay in the room if they wanted. The researcher helped participants feel comfortable by making small talk about school and hobbies before starting the study, following guidelines suggested by prior work on addressing power imbalances between adult researchers and child participants \cite{druin_role_2002}.} Participants then completed two design activities, described below. We concluded workshops with an open-ended discussion with participants about AR apps they had used, what they enjoyed about these apps, and what else they would like to be able to do with them. 

We designed two probes to elicit scenarios that participants would find engaging and fun. \chadded{Our aim with these probes was to release participants' imaginations and to encourage speculation on the potential of AR. To achieve this, we included both a physical and digital probe. We included a physical probe to encourage participants to brainstorm a wide range of scenarios without limiting their ideas to what was technically feasible, as has been done in prior co-design workshops with children \cite{liaqat2022hint, xie2012connecting}}
The first probe included physical toy animals and typical design workshop tools like playdough, markers, pen, and paper. The toy animals functioned as characters, which participants used to brainstorm scenarios. \chdeleted{This probe was designed to encourage open brainstorming without the limitations of technical feasibility.} With the second probe, we aimed to focus idea generation on scenarios more directly grounded in AR and potential tools to create in AR. To support this aim, we developed a rudimentary mobile AR app where users could manipulate an AR character. We selected a block-based approach for manipulating the AR character, motivated by Scratch's effectiveness in eliciting self-expression through a block-based approach \cite{resnick2009scratch}. \chadded{Before introducing the prototype, the researcher first asked the participants if they knew what AR was. For participants who were unfamiliar, AR was described to them as "magic glasses that show you fun things that aren't really there, like cartoons or treasure maps, on top of what you're actually seeing."} We discuss the two probes, the physical design activity and the AR prototype, in detail below. 

\subsubsection{Physical Design Activity}
In the physical design activity, we intended for participants to think about how AR elements could interact with the physical world, and with other AR elements, using plastic toy animals as characters for the basis of these interactions. We placed the plastic animals around the lab space, with each animal placed in a different ``environment'' where young people would typically be on a regular basis (home, park, or school), indicated by an image. We asked participants to walk around and describe what they think should happen when someone (e.g., someone in the workshop) approaches an animal. See Figure \ref{fig:setup} for visuals of our setup. We told participants that their characters were interested in meeting people and having fun. We designed this activity to emulate the experience of interacting with AR characters in the wild, tied to specific locations, such as in popular AR games like Pokémon GO. We omitted discussion of technology from this activity to encourage participants to think beyond what currently exists in AR tools, as has been done in other design activities with children \cite{iacucci2000move}. To encourage brainstorming, we asked questions such as ``What does your character do when it meets someone,'' ``What is your character doing when you are not there,'' ``What does your character do at night?''. Participants wrote their ideas on sticky notes throughout the workshop.

\subsubsection{AR Prototype}
\begin{figure}  
    \includegraphics[scale=0.45]{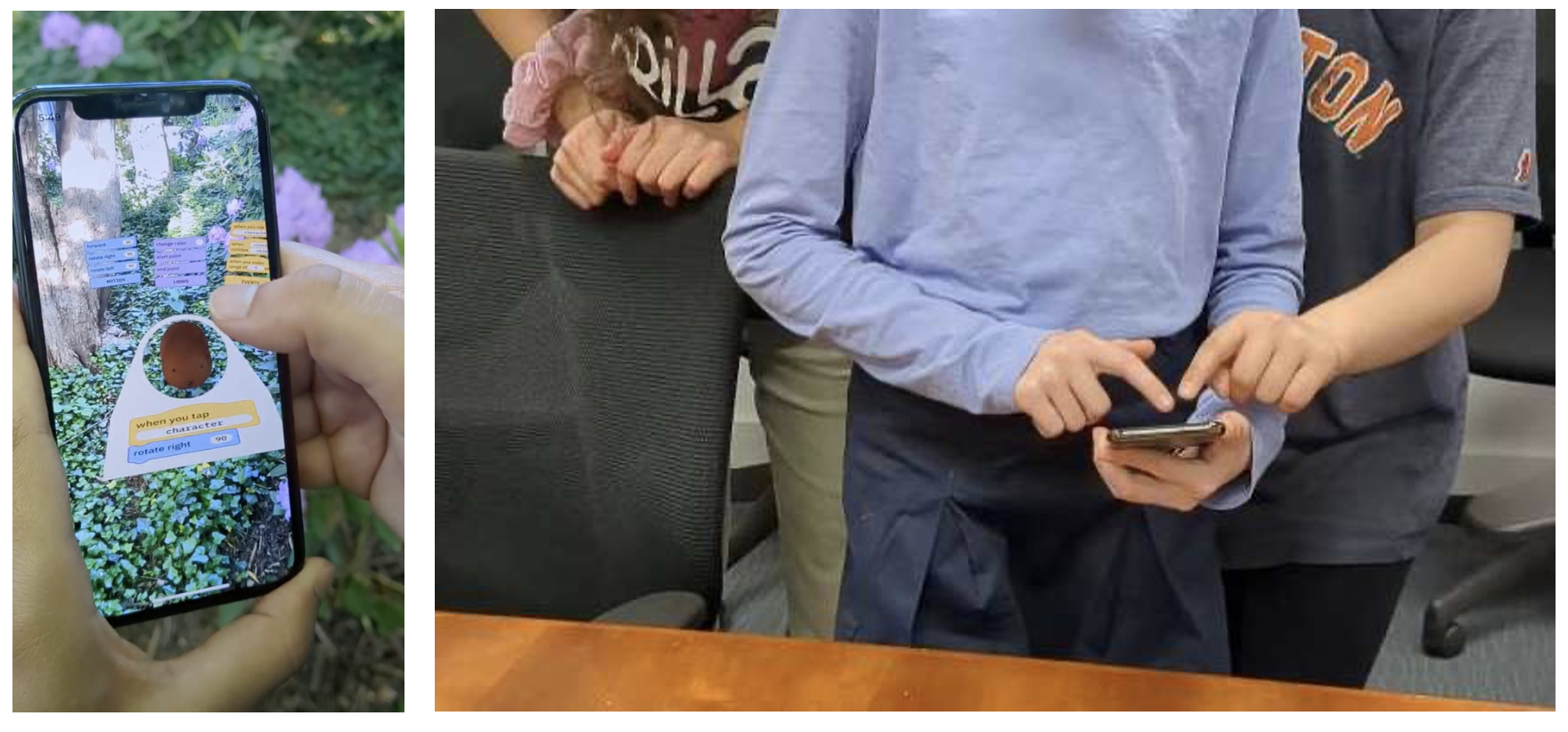}
    \caption{A screenshot of the AR prototype (left). The blocks can be dragged and dropped onto the canvas below the character to create a program. Workshop participants using the AR prototype (right)}
    \label{fig:ar_prototype}
\end{figure}

To focus participants' ideas on scenarios that are more grounded in AR, we used an AR app with basic features inspired by Logo and Scratch. In this app, users could manipulate a 3D animal character (a capybara) through blocks of code with functionality such as moving forward and rotating. Participants used this tool and then brainstormed additional AR scenarios they would like to create. Figure \ref{fig:ar_prototype} shows a screenshot of the prototype. We asked the same prompting questions as the physical probe activity, but in this activity, we encouraged participants to think about how the digital character would interact with the physical characters and other users. 

\subsection{Data and Analysis}
Data collected consisted of audio recordings and their transcriptions, researcher notes, and artifacts created by participants. The four workshops conducted in Spanish were translated into English by one of the researchers. To analyze the data, we performed thematic analysis \cite{clarke2015thematic}. Two of the authors coded two transcripts separately and then discussed the codes together. Codes covered details of the scenarios participants created, such as "Characters learn from each other" and "Participant wants character to be more active during the day". Once this initial codebook was developed, they coded the remaining six transcripts separately and then met again to discuss and resolve disagreements. The two authors grouped similar codes to create themes, which we present in the next section. We did not code for IRR per McDonald and colleagues' recommendation for working inductively \cite{mcdonald2019reliability}.   

\section{Findings}

We present the scenarios, games, activities, and other features participants proposed in the workshop sessions. Scenarios were elicited from both the physical and digital design activities. From the thematic analysis, we created four themes encompassing the creative, social, and technical aspects of the ideas proposed by participants.


\subsection{Participants \chadded{explored different ways to immerse their AR character in the physical world}}

Participants from all eight workshops \chadded{brainstormed ways in which they could immerse their AR character in the physical world. They thought about the technical feasibility of their suggestions by reflecting on how}  the time, location, and camera features, all available in smartphones, \chadded{could be leveraged} to create AR scenarios that interacted with the physical world. Participants \chadded{suggested using} the time, e.g., date or hour of the day, to influence virtual characters' behavior based on routines associated with different times (e.g., sleeping at night). They used location in one of two ways: reacting to new locations and interacting with physical elements. They suggested using the camera features to bring physical objects into AR. \chadded{These ideas indicate a desire to craft AR characters that behave as if they exist in their physical surroundings.} Below, we provide examples of these scenarios that leverage smartphone capabilities: 

\begin{enumerate}
\item \textit{Responding to new locations.} Participants express a desire to control the virtual character's behavior based on where the characters are located, e.g., home or school.
``Maybe interact more with the scenery. [...] Like, it should know where it is. For example, identify what's a school and what's an open landscape. And in the case of an open landscape, perhaps it could explore or interact more immersively with its surroundings.'' --P15AG4.  Similarly,  another participant said: ``Maybe there could be different tasks in different parts of the house, for example, in the kitchen preparing food'' --P16AG4.
 
\item \textit{Interacting with physical elements.} Participants mentioned wanting to 
define spaces, such as rooms in a house, or have boundaries, such as the character stopping when it reaches a body of water. The radius can also be defined relative to the user, such as limiting the character to staying within one kilometer from the user. Or, if the user is at school, the character must also remain within the school. As described by P4US2: ``For phones that have an ultra-wide camera, I feel like it would be really cool if you were able to, like, take advantage of that... maybe you could, like, scan a table and then that would be like the character's play space. It couldn't, like, leave.''  P10AG5 suggested a similar idea: ``I'd also like it if you could draw something on a piece of paper. It detects it with the camera and puts it in the video game. I draw something, put it there, and it detects it in the game. And it can play with it.'' Participants enjoyed pretending that their characters existed in the physical world, as can be seen in Figure \ref{fig:interaction}. When asked where her AR character was, P6US3 pointed to a spot on the ground rather than her phone screen.
    
    \begin{figure}  
    \includegraphics[scale=0.5]{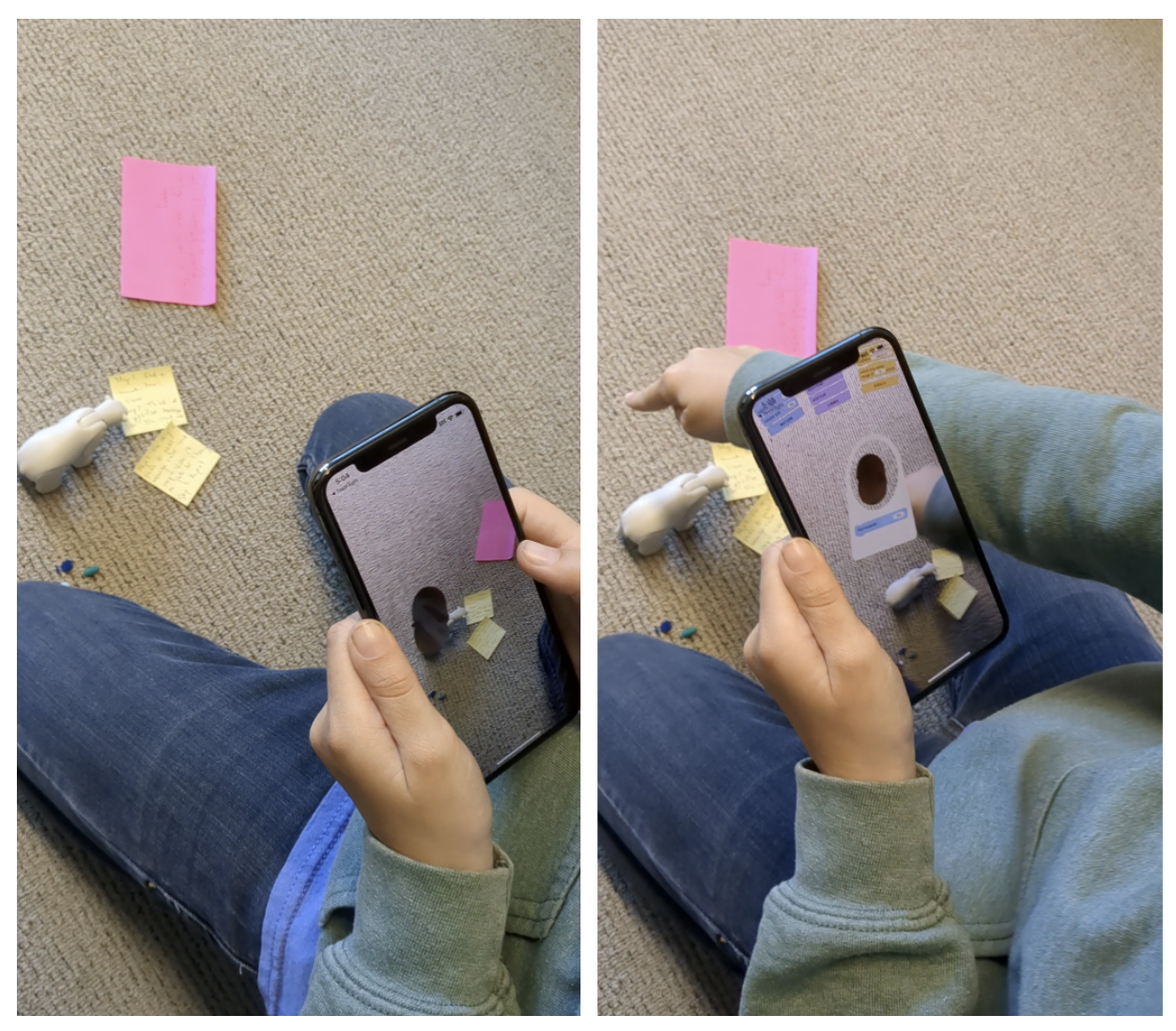}
    \caption{P6US3 manipulates the AR character to walk over the polar bear figurine. (left). When asked by the researcher where the AR character is, she points to a physical spot on the ground rather than the phone screen. (right)}
    \label{fig:interaction}
\end{figure}

\item \textit{Time-based behaviors.}The character's greetings can be time-based, such as ``Good morning'' or ``Happy Easter.'' As described by P5US2: "You can have it so that it looks at your phone's time, and it will have a little sun and moon cycle, day and night cycle. So when it's nighttime, the [character] has
a little sleeping animation." P11AG1 suggested: "If it's sunny, it should be sunny. And when it's nighttime, for example, it should be nighttime. And so on."

\item \textit{Other smartphone data.} Participants mentioned wanting to take advantage of additional information available on smartphones, such as data on other apps the user was running: ``For instance, playing [a game on your phone], it could look at the screen and recognize the score... Like, there's an interaction with what the person is playing. So, when you're playing, and it shows on the screen, 'you won,' things like that, the pet could react to that'' --P16AG4.
\end{enumerate}

From these scenarios, we found that \textbf{participants wanted to use their smartphones' capabilities to create more immersive AR experiences when interacting with the physical world}. Participants enjoyed creating scenarios that grounded their AR character in the real world, and they sought out ways to have their character engage with, and respond to, changing world conditions that added elements of realism to interactions.    

\subsection{Participants wanted to make location-dependent scenarios with other users, but these  scenarios can be asynchronous} \label{sec:location-dependent}

In four of the workshops, participants suggested creating location-dependent and asynchronous scenarios, i.e.,  people must be physically in the same spot but not necessarily at the same time. Some of these asynchronous scenarios were games that could be played over a long period of time and typically involved turn-taking or "roles" assigned to users. Many of these games have non-digital parallels, \chadded{like the children's game Tag,} that participants adapted and enhanced in the context of AR - \chadded{for example, by making the AR character a player that can also run around and tag people.} Interestingly, participants \chadded{emphasized that such games would be socially engaging and rewarding,} even though the interaction was not happening synchronously. \chadded{This sense of social presence comes from viewing and responding to actions their friends had made.}  Players would communicate through the app, and each player could respond and complete their portion of the game in their own time. \chdeleted{Participants gravitated towards asynchronous even when synchronous versions of the game were possible.} As described by P4US2: ``you could create challenges and you could post them on an online thing and then you could play other people's challenges.''

Examples of scenarios participants envisioned that were location-dependent and asynchronous included:

\begin{enumerate}
    \item \textit{Finding an Object.} In this game, a user drops an object such as a pin, star, or flower. They then invite their friends to program their character to walk to the pin.   
    \item \textit{The Hot Cold Game}. In this game, a user leaves an object somewhere within a defined radius (e.g., school). The user then sends a challenge to their friends. The friends must program their character to walk around, and the app tells them if they are getting warmer or colder. They have to change their code based on this information. This game can be played with multiple friends as a challenge, with the user sending a message to indicate when the challenge starts.  
    
    \item \textit{Tag.} In Tag, one AR character in the friend group is “it.” If you spot the "it" character in the wild, you have to run away. Participants proposed an asynchronous version of this game by having it take place over days or weeks within a friend group. Figure \ref{fig:tag} shows what the tagging would look like. P7US3 manipulated the AR character to walk to P6US3, who ran away as the character approached her. A variant of this game participants suggested is Hot Potato. In Hot Potato, characters have to pass around an object within a certain time limit. This game is a cross between asynchronous and synchronous as it requires two users to meet at the same time, but the game can stretch over a longer period of time. 

    \begin{figure}  
    \includegraphics[scale=0.4]{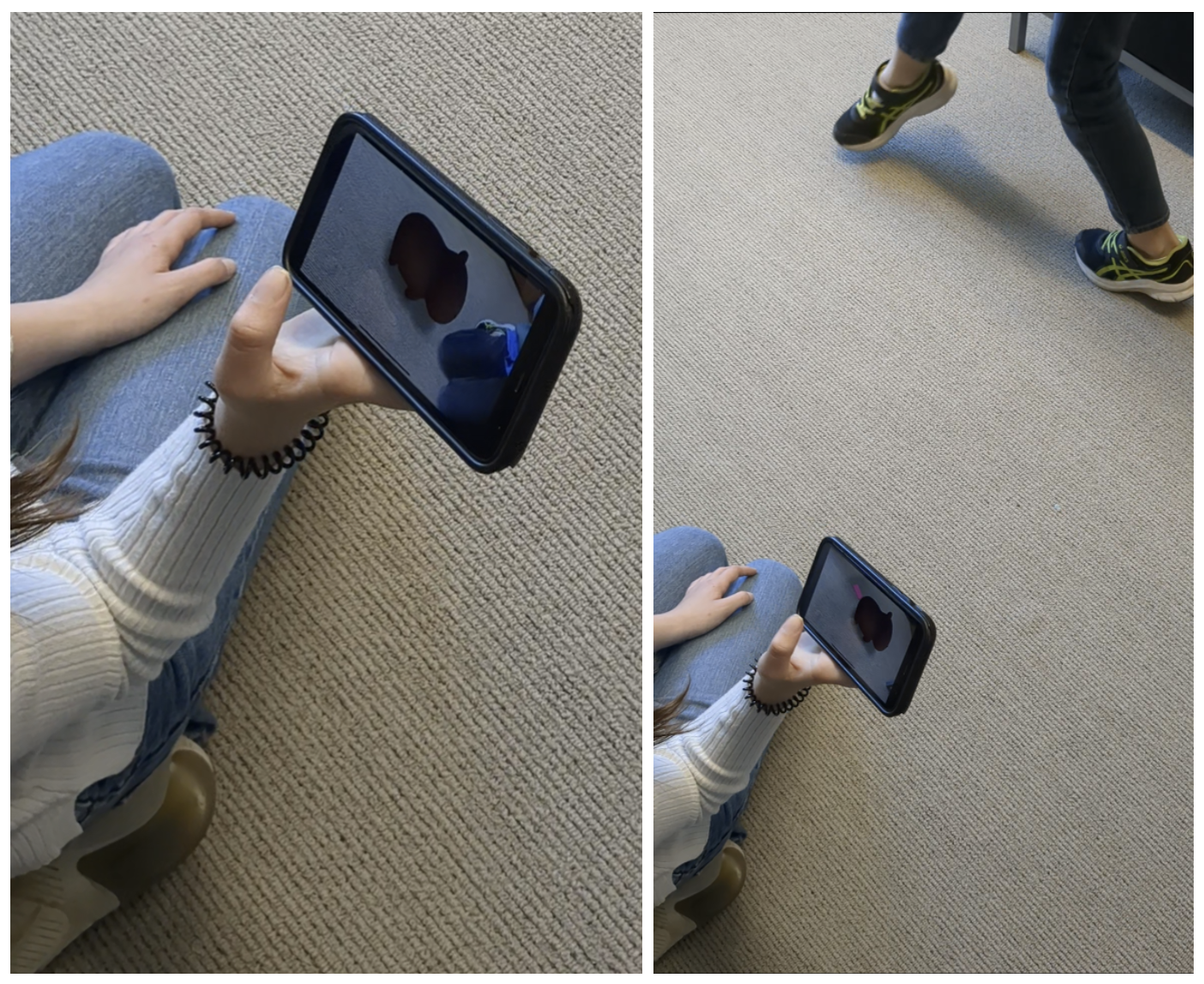}
    \caption{P7US3 manipulated the AR character to walk towards P6US3 (left). When P7US3 informed P6US3 that she is about to be tagged, P6US3 runs away (right).}
    \label{fig:tag}
\end{figure}

    \item \textit{Leaving messages}. Participants discussed how communication between users would work. Chat messages were suggested, as was the idea of leaving messages. As seen in Figure \ref{fig:messages}, a character can leave a rock with a painted message for other users to discover. 
    \begin{figure}  
    \includegraphics[scale=0.05]{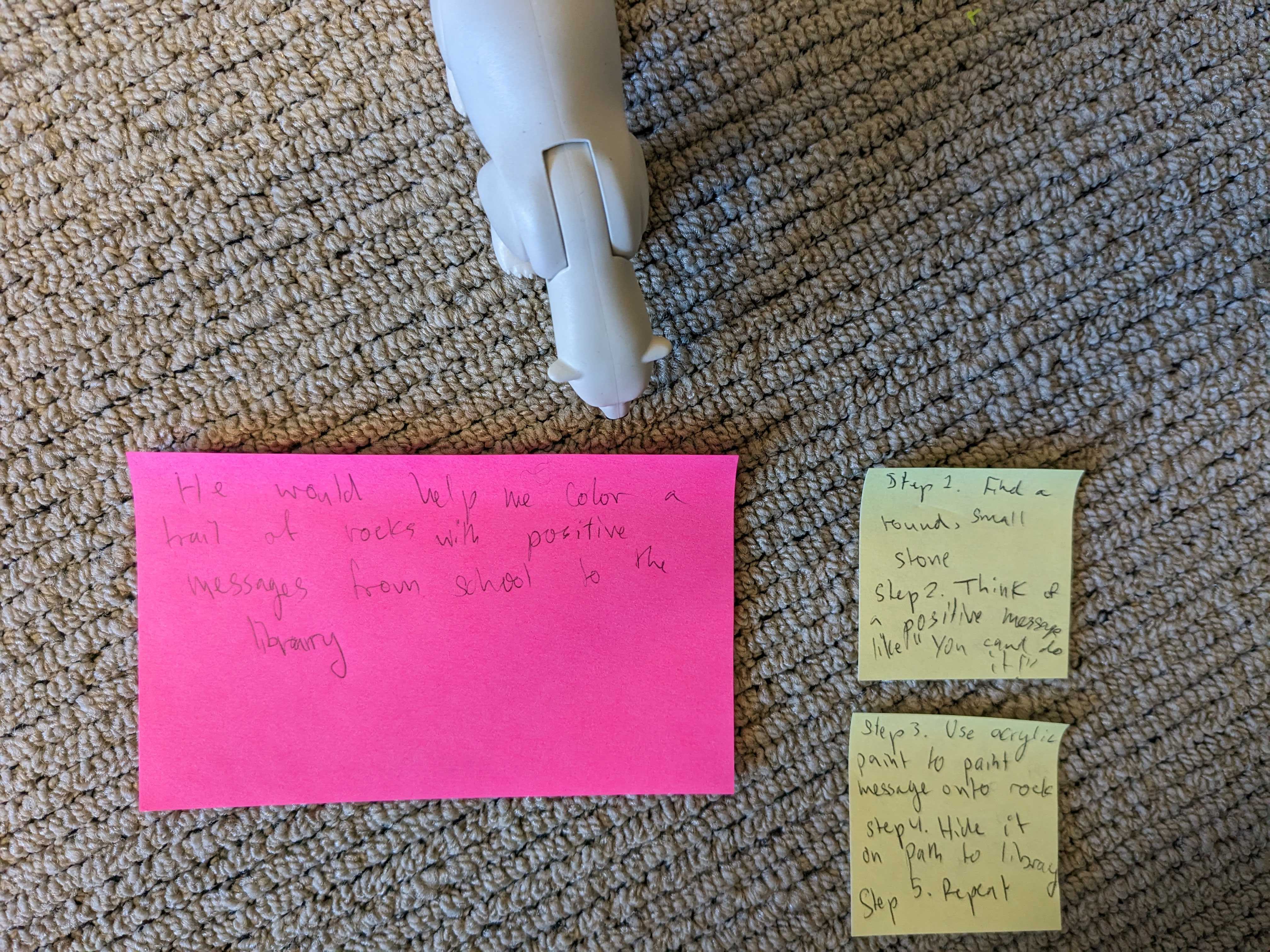}
    \caption{Participants from US3 suggested that characters should be able to leave inspirational messages for other users to discover. The large pink sticky note on the left reads: "He [the polar bear] would help me color a trail of rocks with positive messages from school to library". The smaller yellow sticky notes on the right list the steps: "1) Find a small, round stone. 2) Think of a positive message like 'You can do it', 3) Use acrylic paint to paint message onto a rock, 4) Hide it on the path to the library, 5) Repeat"}
    \label{fig:messages}
\end{figure}
    
\end{enumerate}

From these scenarios, we suggest that participants were interested in designing asynchronous games and interactions. By placing AR elements in locations where other users could find them, participants created games that were fun without requiring users to be present at the same time. These scenarios let participants leave behind a "trace" of their presence, such as leaving messages on rocks for others to find (physical-world parallels may be carving names into a tree or leaving a lock on a fence). Participants also suggested synchronous scenarios (e.g., building castles together, synchronous tag, coordinating meetups by arranging a location and time).  

Participants found these multiplayer scenarios that took place over a longer period of time to be socially engaging. Designing to support the development of these asynchronous scenarios means that \textbf{participants were interested in designing games that involve turn-taking and cooperative features between multiple players}. \chadded{This collaborative building on top of other's contributions distinguishes the scenarios participants made from mainstream AR apps like Pokemon Go, where players can meet each other and interact, but there is no collaborative creation.} 

\subsection{Participants wanted \chadded{to foster a sense of togetherness and friendship between AR characters through meetings in physical space} }

Participants from seven workshops thought about designing for scenarios where characters could interact with each other \chadded{and build relationships through meeting one another and exchanging knowledge.} For these scenarios, we found that participants were interested in sharing and learning from other users' characters. We describe some of these scenarios below: 
\begin{enumerate}
    \item When a user's character runs into other AR characters, there should be options for seeing the other character's code, copying it and modifying it. Participants from US2 described what this might look like: ``Say they built a castle, you could add another castle next to it or something. And you could save it, and it would be your own'' (P4US2). P5US2 added, ``And it would say remix of blank.'' Both participants in this workshop used Scratch, which has options for remixing code. These scenarios that required collaboration reflect participant behavior during the workshops. We found that participants enjoyed collaborating when creating, and we observed that participants worked together to craft scenarios, as can be seen in Figure \ref{fig:collab}.

    \begin{figure}  
    \includegraphics[scale=0.1]{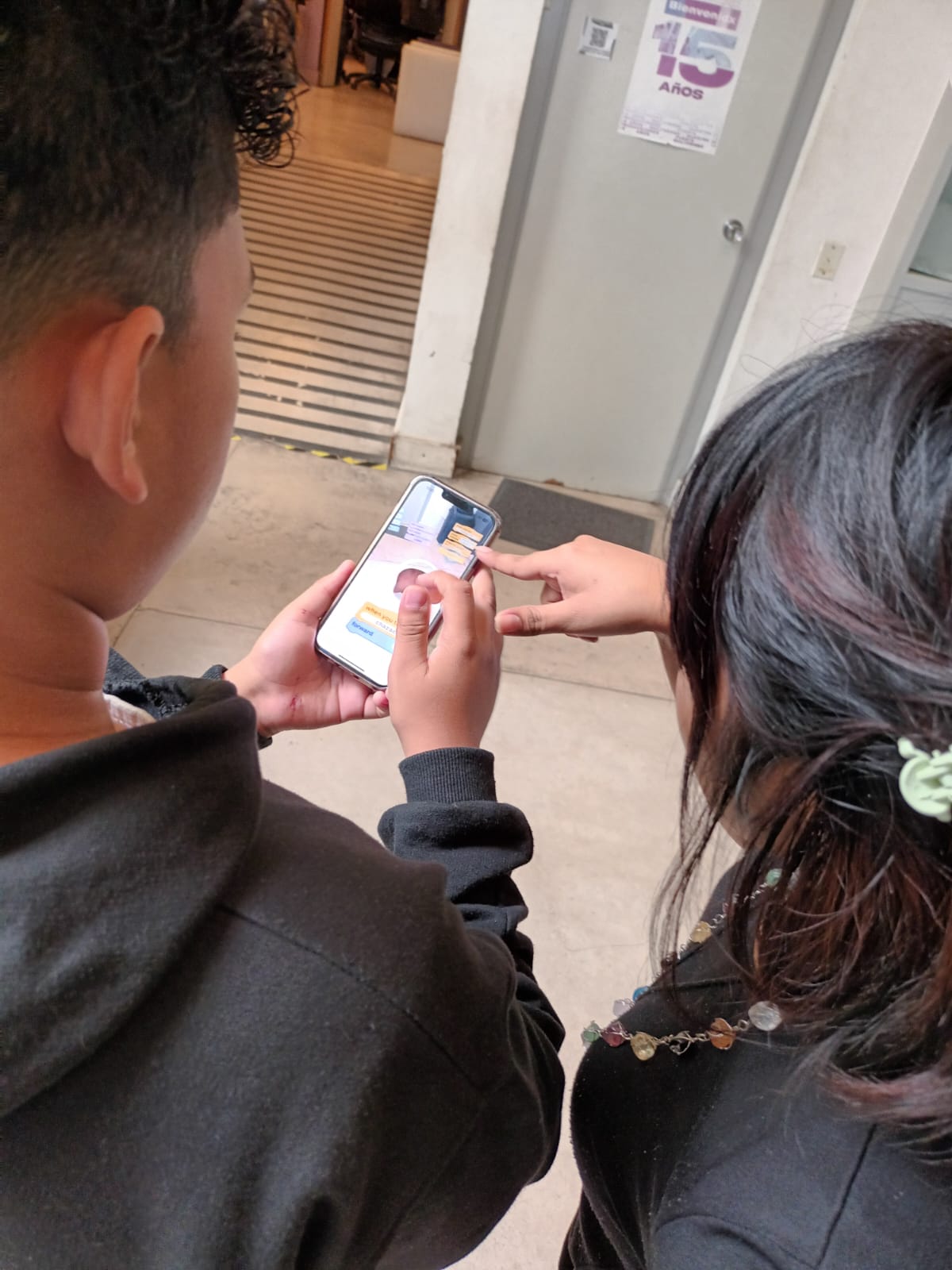}
    \caption{Participants from AG1 worked together to create a scenario on the AR prototype app.}
    \label{fig:collab}
\end{figure}

    \item If a user's character runs into another character in the wild, the character should have different reactions to a friend, e.g., an offer to do a challenge, versus to a stranger (introduce yourself). P10AG1 said: "They could have options like in The Sims, you know? When you interact with someone else, it allows you, I don't know, to give a hug. For example. I don't know, a little ball could appear, and they could start playing."
\end{enumerate}

P12AG2 described a multiplayer scenario: \begin{quote}
"You have your [character] there, and of course, on the other side, she has the penguin, and I can see her penguin. Like, you see both of them at the same time. She sees mine, and I can see hers.

And they could appear in the app like Pokémon. Like, you pass by and see her thing, but without being linked by a code or something. And they can see each other because it's part of the app.

And it could be that you can program them to be together, like walking hand in hand, as I said. Like becoming friends. And every time you see them online, you can talk to them or something."
\end{quote}


Participants' scenarios implied they were interested in fostering a sense of togetherness and representations of friendships with other characters. These ideas manifested in the scenarios where participants wanted to engage with other characters, learn about them, and build relationships over time. The scenarios created by participants suggest that \textbf{they are interested in programming conditional scenarios that react to or build on other users' behaviors}.

\subsection{Participants wanted information asymmetry between characters and users to create discovery opportunities}

We observed that participants in all eight workshops interacted with their characters \chadded{and imagined that their characters were responding}. Participants would try to teach their characters, and they also brainstormed scenarios where they could learn from their characters. This discovery exchange is made possible when the user has information the character does not (e.g., "proper" ways for a pet-like character to behave), and vice-versa (e.g., the character has information about their surroundings). We describe some of these learning moments below: 
\begin{enumerate}
    \item The user raises the character as if it were a pet. This includes helping it develop a personality such as likes, dislikes, reactions, etc. P1US1 said: ``You're like coding it to interact with it. It's like a virtual pet.'' P2US1 agreed, expanding on the idea by suggesting that they could set personalized requirements for their character, such as the amount of exercise or food required. The character would then respond when interacted with based on these requirements (e.g., goes from sad to happy when fed): ``Your animal, like, it, has like a certain amount of requirements that you need to code for them. Like a certain amount of exercise... Maybe they eat something, and you need to boost their mood.''

    \item The character can use recommendation algorithms to provide curated suggestions to participants. P15AG3 suggested: "By considering their hobbies and talents, the character could provide tips, tutorials, or methods for crafting, cooking, or knitting techniques...like an assistant that helps them out." P17AG4 had a similar idea: ``My animal would give recommendations, since I see here that he (P16AG4) likes a lot of anime, manga, books, and so on. A recommendation, kind of like the algorithm that YouTube, Netflix use, you know?'' With these scenarios, participants were implying that they would like to design characters that had knowledge they did not. This would create opportunities for thoughtful interactions, such as making suggestions to a friend based on their interests. 
 
    \item When visiting a new location, the character can share interesting facts about the area, such as identifying plants or wildlife. P14AG3 described: ``It could visit some places that might be difficult for a human to visit. If it can discover something new or if I could see what it's seeing at that moment, that would be great. Or if it could share interesting information about the place or something like that.'' P15AG3 from the same workshop built off the idea: ``I imagine it could take photos. In places you can't go, it could take photos.'' With these scenarios, participants reflected on how both they and their character may behave in different ways in new locations. New locations could provide opportunities for the user to learn something from the character, who has access to information (i.e., smartphone data) the participant does not. Another related idea participants suggested was that characters could also teach each other. P14AG3 said: ``Both characters could teach different things. Maybe one knows how to do something that the other doesn't. So they could teach each other.''

\end{enumerate}

Participants viewed their characters as a lens through which they could gather interesting information about the physical world. They also enjoyed the idea of their characters' behavior changing over time as they learned. From these scenarios, we suggest that \textbf{participants were interested in characters that can teach users and, conversely, can respond to new information by changing their own behavior.}

\section{Discussion}
AR provides opportunities for creating interactions between the physical and digital worlds. In our study, we prompted young people to explore what these possibilities could look like \chadded{to identify what their goals are for an AR authoring tool.} 
We identified four general themes that touched on the social, creative, and educational goals of young people. These themes point to the unique affordances of AR, as the blending of the digital and physical sets AR technology apart from other modalities \chadded{like VR}. In Table \ref{tab:discussionsummary}, we summarize our themes from our findings, \chadded{identify the value or goal of young people revealed in the findings} \chdeleted{present the tensions we observed}, and list the \chdeleted{design} opportunities for designing AR technology \chadded{that supports the goals young people have for creating with AR.}


  \begin{table*}[h]
  \begin{tabular}{p{0.5cm}p{4cm}p{4cm}p{4cm}}
    \toprule
    \textbf {} & \textbf{Study Findings} & \textbf{\chadded{Underlying Value or Goal}} & \textbf{Opportunity for Design} \\
    \midrule
    \ref{sec:discussion1} & \chreplaced{Participants explored different ways to immerse their AR character in the physical world using technical features and their imaginations.}{Participants wanted to ground their characters through interactions with the physical
world via their smartphone’s location, time, and camera features.} & \chreplaced {Young people want to make realistic experiences that represent the narratives they imagine.} {Limitations of AR technology do not allow for creation of highly realistic scenarios.} & \chadded{Blend realism with imagination to fill in the gaps of young people's scenarios.}   \\
    \midrule
    \ref{sec:discussion2} &  Participants wanted to make location-dependent scenarios with other users, but these scenarios can be
asynchronous. & \chadded{Even when they perform an action with no one else around,} young people want their contributions to be \chadded{recognized}. & \chadded{Explore how traces can be represented in AR to make contributions visible.} \\
    \midrule
   \ref{sec:discussion3} &  \chreplaced{Participants wanted to foster a sense of togetherness and friendship between AR characters through
meetings in physical space}{Participants wanted to interact with other users’ characters to learn, play, and socialize.} & \chreplaced{Young people want to communicate feelings of social connection to their peers.}{Transparency and sharing in AR environments must be balanced with young people's autonomy and social norms.} & \chreplaced{Consider how AR artifacts can serve as invitations to connect} {Leverage AR to craft moments of discovery in physical and temporal space that mimic social conventions.} \\
    \midrule
   \ref{sec:discussion4} & Participants wanted information asymmetry between characters and users to create learning
opportunities. & Young people \chreplaced {know what they are interested in learning and how they would like to pursue the information they want to learn.} {have diverse goals and are interested in different information.}  & \chadded{Enable discovery of} information that young people care about.  \\
 \bottomrule
  \end{tabular}
  \caption{Summary of our study findings, the \chreplaced{the underlying value or goal the finding suggests}{tensions we observed}, and opportunities for designing AR experiences in line with young people's goals.}
  \label{tab:discussionsummary}
\end{table*}

\subsection{\chadded{Blend realism with imagination to
fill in the gaps of their scenarios}} \label{sec:discussion1}
\chadded{We suggest that designers of AR authoring tools consider how young people's imaginative narratives can be preserved in the experiences they create.}
In our study, participants used their characters as a lens for exploring the physical world, and created narratives to accompany these scenarios.  They \chdeleted{interest in storytelling as they} enriched the scenarios they created with character backstories (e.g., some characters are friends while others are not, which changes their behavior when interacting). \chadded{This imaginative, playful creation distinguishes the young participants in our study from prior work with adults. Even with AR authoring tools intended for creating realistic experiences, adult participants did not craft oral narratives to accompany their experiences \cite{gasques2019pintar, leiva2020pronto}. However, this narrative-building behaviour has been observed in children using both AR and non-AR tools. In an AR app for making treasure hunts, children told stories about the maps they made \cite{oberhuber2017augmented}, while in a non-AR online game, children created complex, multi-linear stories to personalize their digital artifacts  \cite{garzotto2006hyperstories}. We suggest that AR authoring tools for young people should explore how imaginative elements can be incorporated into scenarios to engage users. For example, designers might draw inspiration from audio recording features in non-AR tools like StoryTapestry~\cite{liaqat2023exploring}, in which young people might audio record their thoughts, questions, and narratives as they work.} 

We encourage designers of AR authoring tools to think more broadly about what authorship means to young people. We found that young people enjoy using their imaginations when creating AR experiences. Blending imagination into these experiences can help create more accessible AR authoring tools, as users who are not interested in creating complex scenarios (e.g., younger users) can incorporate imaginative elements into digital creations to create rich, personalized experiences. In summary, beyond the digital artifacts they make, the imaginative narratives that underlie young people's creations are also important intellectual contributions that should be preserved in an AR authoring tool.

\chdeleted{and creating in digital spaces has been found to enrich the stories they build, such as in a tablet-based AR app for making treasure hunts  \cite{oberhuber2017augmented} or constructing digital storybooks to accompany oral stories \cite{liaqat2023exploring}.}

\chdeleted{Therefore, we suggest that AR tools offer young people the option to configure the behavior of AR artifacts, to create scenarios where these artifacts can interact with the physical world in realistic ways, and to encourage them to build stories around their creations. These affordances can broaden the ways in which young people can creatively express themselves in digital spaces through the creation of AR artifacts that young people can then craft stories around. From our study, we found opportunities for extending how we, as designers, think about the personalization of digital characters. Beyond modifying just a character's physical appearance (e.g., clothes, accessories, etc.), designers of AR authoring tools can offer a richer array of configurations by allowing users to create characters that respond situationally to the physical environment. In this way, young people can craft characters that initiate interactions in response to their changing environments, creating moments of fun and surprise.}    

\chdeleted{Participants took the behavior configuration of their characters a step further by taking advantage of location time, camera, and other features in smartphones to program their character's behavior. With access to this information, it is possible to offer a broader range of creations for young people, allowing for greater creative expression.}


\subsection {\chreplaced{Explore opportunities for how traces can be represented in AR to make contributions visible}{Turn-taking games with multiplayer functionality should allow for leaving traces as evidence of contribution}} \label{sec:discussion2}

\chadded{Visibility of contributions was important to participants, and we observed how this value manifested as our participants created collaborative scenarios.} \chdeleted{Many scenarios involved the multiplayer functionality. This was due in part to the design of our workshops, as we asked participants what they would like to do with other characters and users they encounter in the wild.} \chadded{Some of the collaborative scenarios participants created} included location-dependent and asynchronous aspects \chdeleted{of these interactions.} The scenarios that they brainstormed mostly depended on turn-taking interactions in a specific location, such as \textit{Finding an Object} and \textit{Tag}, as suggested in~\ref{sec:location-dependent}, and many of these scenarios involved the idea of leaving behind traces that other users could identify. \chadded{The process of creating these asynchronous scenarios revealed that young people value the visibility of contributions, even when an action is performed alone.} 

\chadded{In our study, we found that young people want to leverage location-based scenarios to establish ownership and also to create surprises for their friends. Prior work has shown how children can leverage location in AR to influence the behavior of other users, such as when creating trails that lead to a particular store while intentionally avoiding others \cite{lunding2022exposar}. }. Prior work has also identified shared attention and physical manipulation as aspects of AR that influence how collaboration unfolds \cite{im2021draw2code, wu2023mr}. In our study, we observed how shared location additionally influences collaboration by encouraging asynchronous interactions. Therefore, we suggest that AR authoring tools for young people should allow users to claim ownership over their contributions, and that these traces should be visible so that other users can discover them. To boost young people's surprises and joy from such asynchronous location-dependent scenarios, we propose that designers \chadded{explore different ways in which traces can be represented in AR. For example, this could be basic but engaging asynchronous communication features such as leaving messages to preserve ownership. More immersive traces could bring together physical and digital elements, such as writing a message on a physical rock or tree.}

The HCI community has long been exploring how to design location-dependent experiences to benefit social interactions~\cite{paasovaara2017understanding, liu2022understanding, 10.1145/3025453.3025871, 10.1145/3025453.3025761}. One common point that these studies raise is that location-dependent experiences are suitable for strong social ties such as friends, family, and partners~\cite{brown1987social}. Building on this strand of work, our work found that young people were especially interested in the asynchronous mode in location-dependent scenarios as a way of social connection. Asynchronous interactions over a period of time in one specific location could potentially reward them with delight in their friends' reactions \chadded{after some time has passed}. 


\chdeleted{
consider adding complex scenarios while still preserving basic but engaging asynchronous communication features such as leaving messages to preserve ownership. To make the scenario more complex, designers can consider providing multi-step challenges for users to program. For example, in the example of \textit{Finding an Object} proposed by participants, users first drop an object and invite their friends to program their character to walk to the object. To make this process more challenging and engaging, apps can suggest the user who initiates the invitation options add multiple locations and problem-solving tasks before the invitee finds the object. In addition, preserving basic asynchronous communication methods enriches the layers of communication difficulties and challenge levels, and for leaving traces of contributions which we found is important for young people.}

\chdeleted{It should still be noted that any location-based games should take safety into account, especially in public spaces. As suggested by prior work~\cite{ekman2005designing}, apps can consider adding audio cues to remind users of their safety when playing outside (e.g., in traffic). Besides audio cues, we also suggest adding some limitations in terms of where users can possibly drop their items for their friends to find. For example, they cannot drop an item in the middle of a crossroad.}

\subsection{\chreplaced{Consider how AR artifacts can serve as invitations to connect}{Autonomy can be preserved when creations are shared by using spatial and temporal features in AR to mimic social conventions} } \label{sec:discussion3}
Young people were interested in the social aspects of the experiences they made with other users, \chadded{and the artifacts they created served as symbols of connection (e.g., building castles side by side with friends)}. This interest in seeking social interactions reflects what we know about young people, who seek to develop unique identities while pursuing meaningful social relationships with their peers \cite{tomova_importance_2021}. We know that they draw heavily on the socialization patterns they observe to form their own identities and a sense of belonging \cite{coll_immigrant_2009}. \chadded{Young people's social goals sets them apart from adults. We see this in AR authoring tools for adults, which focus on creating realistic, complex experiences. Prior work has identified that children enjoy predicting how their peers may react to AR scenarios they create and in designing experiences to manipulate their peers' behavior \cite{lunding2022exposar}.} In our study, we observed two steps in participants' social interactions. First, participants engaged with the character (greeting it or introducing themselves if it was a stranger). After this, they tried to learn more about the scenario the other character was running. 

For designers of AR authoring tools, we suggest \chadded{considering how AR artifacts can serve as invitations to connect as well as visual symbols for social connection.} \chdeleted{several opportunities to support these social interactions for young people.} This could include features that support young people in sharing, copying, modifying, and asking questions about the scenarios they create \chadded{to encourage communication}. This is similar to the idea of remixing code in Scratch \cite{resnick2009scratch}, where users can build on other users' projects. However, compared to desktop-based creations, AR \chreplaced{offers additional dimensions} {presents several opportunities} for enriching these social interactions \chadded{through spatial and temporal features}. First, \chadded{the spatial dimension of AR opens up opportunity for} discovery. Young people can discover artifacts (e.g., a castle) left by other users, and they can run into other characters in the wild. We found that young people enjoyed these moments of "stumbling" upon these interaction opportunities in the wild, \chadded{and that these discoveries opened the door for social connection, such as by collaboratively building on a peer's creation so that both are authors of the experience}. Second, the concept of time was highlighted by participants. Social interactions can be made more dimensional when they unfold over a period of time, such as going from strangers to friends as you repeatedly come across the same user. We suggest that designers consider leveraging these concepts of location and time interplay with concepts of trust, friendship, idea ownership, and relationship building to allow young people a richer social experience in \chadded{an AR authoring tool}. 

\chdeleted{
The young people in our study felt that it was important to have transparency in the environment, in the form of being able to see how others had built their scenarios. This transparency, after seeking permission from the user, was important for creating engaging experiences and for learning.}


\subsection{Enable discovery of information that young people care about} \label{sec:discussion4}
\chadded{}
Participants designed scenarios where their characters could teach them something new about their surroundings, while participants would teach their characters to behave and react to their surroundings and people. \chdeleted{This reciprocal behavior we observed relates to our earlier finding (\ref{sec:discussion1}) on young people's interest in personalizing their characters.} Researchers have leveraged information asymmetry between users to invite social interaction and conversation in an AR escape room game \cite{knoll2023arctic}, a web-based storytelling tool for families \cite{liaqat2022hint}, and digital games with mixed-sighted pairs \cite{gonccalves2021exploring}. In our study, participants proposed intentional information asymmetry between themselves and their \chreplaced {characters}{non-human AR artifacts}. Young people \chdeleted{viewed their characters as semi-autonomous, so while they} wanted to program their character's behavior to a degree, and they also wanted their characters to initiate interaction with them. \chadded{We suggest that designers of AR authoring tools for young people consider how AR artifacts can be used as a lens for learning about the physical world. For instance, users can} create characters that change over time as they learn new information. Location can be leveraged by designers to let young people create characters that can provide location-specific information and feedback (e.g., an interesting fact about a nearby landmark or a suggestion for where to eat). As suggested by participants, one way to implement this would be to use recommendation algorithms for characters to make suggestions and share information.

\chdeleted{As with Finding  \ref{sec:discussion3}, location and time play key roles in fostering these reciprocal user-character interactions.}

\section{Limitations and Future Work}
Our design workshops generated scenarios that young people are interested in creating in AR. While we found a variety of scenarios that can be sources of inspiration for designers, there are several limitations in our work that future research should address. First, we centered scenario generation around a character, based on prior work and popular AR apps, to help participants ground their brainstorming. Follow-up design workshops could broaden the scope to focus on other types of AR artifacts (e.g., building blocks). Second, our participants had prior experience with programming and creation tools such as Scratch, and brought up several ideas related to coding were brought up by participants due to their backgrounds. Future studies should explore the design of AR experiences with young people who have no programming experience, to understand the types of scenarios they would like to make without that prior knowledge. Finally, while we employed design workshops to incorporate young people's voices in the early stage of the development process, future work should pursue additional approaches to further explore the potential of AR in supporting young people's self-expression. This may include prototyping and evaluation of AR tools that are informed by findings from this study. 

\section{Conclusion}
 In our study, we investigated the experiences that young people would like to create in AR. Since many AR apps have been developed without the involvement of young people, we sought to identify scenarios that young people are interested in incorporating into \chadded{an AR authoring tool} through \chadded{user-centered}\chdeleted{a series of} design workshops. We identified four interests, goals, and needs of young people and made four resulting recommendations for designers of AR technology, including supporting:  \chadded{(1) Blending imagination into AR scenarios, (2) Representing traces of contribution in AR, (3) Using AR artifacts as invitations to connect, and (4) Enabling discovery of information.} Overall, these two sets of contributions - empirical insights into young people’s creative goals and design
recommendations for future AR experiences - help set the stage for future investigations into designing AR experiences that empower young people to create.








\bibliographystyle{ACM-Reference-Format}
\bibliography{chi_refs}










\end{document}